\documentclass{nature} 
\usepackage{graphicx}
\usepackage{caption}
\usepackage{lineno}
\usepackage{authblk}
\usepackage{hyperref}
\usepackage{float}
\restylefloat{table}

\usepackage{amsmath,amssymb,amsxtra,amstext,amsfonts,dsfont}
\usepackage{color}
\usepackage[normalem]{ulem} 
\usepackage{multirow}
\usepackage{array}
\usepackage{upgreek}
\usepackage{mathrsfs}

\usepackage[squaren]{SIunits}

\newcommand{\bra}[1]{\ensuremath{\langle{#1}|}}
\newcommand{\ket}[1]{\ensuremath{|{#1}\rangle}}

\newcommand{\abs}[1]{\ensuremath{\left|{#1}\right|}}
\newcommand{\oa}{\omega_a}
\newcommand{\oq}{\omega_q}
\newcommand{\sz}{\sigma_{z}}
\newcommand{\spl}{\sigma^{+}}
\newcommand{\smn}{\sigma^{-}}
\newcommand{\adg}{a^\dagger}
\newcommand{\D}{\Delta}
\newcommand{\e}{\epsilon}
\newcommand{\HJC}{H_{\text{JC}}}
\newcommand{\Dpr}{\Delta^\prime}
\newcommand\mr{\mathrm}

\newcommand{\id}{\mathbb{I}}

\title{Parity measurement in the strong dispersive regime of circuit quantum acoustodynamics}

\author[1, 2, $\dagger$, *]{Uwe von Lüpke}
\author[1, 2, $\dagger$]{Yu Yang}
\author[1, 2, $\dagger$]{Marius Bild}
\author[1, 2]{Laurent Michaud}
\author[1, 2]{Matteo Fadel}
\author[1, 2, **]{Yiwen Chu}

\affil[1]{Department of Physics, ETH Zürich, 8093 Zürich, Switzerland}
\affil[2]{Quantum Center, ETH Zürich, 8093 Zürich, Switzerland}

\affil[*]{vluepkeu@phys.ethz.ch}
\affil[**]{yiwen.chu@phys.ethz.ch}

\affil[$\dagger$]{these authors contributed equally to this work}

\begin{document}
\maketitle
\newline
\begin{abstract}
Mechanical resonators are emerging as an important new platform for quantum science and technologies. A large number of proposals for using them to store, process, and transduce quantum information\cite{Pechal2019, Hann2019, Chamberland2020, Chu2020} motivates the development of increasingly sophisticated techniques for controlling mechanical motion in the quantum regime. By interfacing mechanical resonators with superconducting circuits, circuit quantum acoustodynamics (cQAD) can make a variety of important tools available for manipulating and measuring motional quantum states. Here we demonstrate direct measurements of the phonon number distribution and parity of nonclassical mechanical states. We do this by operating our system in the strong dispersive regime, where a superconducting qubit can be used to spectroscopically resolve phonon Fock states. These measurements are some of the basic building blocks for constructing acoustic quantum memories and processors. Furthermore, our results open the door to performing even more complex quantum algorithms using mechanical systems, such as quantum error correction and multi-mode operations. 
\end{abstract}
\maketitle

The quantum properties of solid-state mechanical objects have now been unequivocally demonstrated through a number of seminal experiments\cite{Chu2018, Bienfait2020, Arrangoiz2019, Sletten2019, Kotler2021, Ockeloen-Korppi2018}. By interfacing mechanical objects with the strong quantum nonlinearity of superconducting qubits and drawing close analogies to the well-developed field of circuit quantum electrodynamics (cQED), cQAD has become a particularly powerful paradigm for creating and studying mechanical quantum states\cite{Chu2018, Bienfait2020, Arrangoiz2019, Sletten2019}. At the same time, mechanical resonators present a set of unique and useful properties that distinguish them from their electromagnetic counterparts used in cQED. For example, acoustic resonators can be more compact, suffer less from crosstalk, and exhibit longer lifetimes \cite{MacCabe2020, Tsaturyan2017} than electromagnetic resonators. The fact that they are massive mechanical objects that exhibit quantum behavior also makes them interesting platforms for investigating a variety of questions in fundamental physics\cite{Gely2021, Pikovski2012}.

A clear and immediate goal for the field is then to use mechanical devices to demonstrate more complex quantum operations. In order to achieve this challenging goal, cQAD allows us to adapt tools and techniques from other fields. A particularly useful set of tools becomes available when a coupled qubit-resonator system reaches the strong dispersive regime, where an excitation in the qubit (resonator) results in a shift of the resonator (qubit) frequency that exceeds the decoherence rates of both systems\cite{Bertet2002, Schuster2007}. The dispersive interaction is nowadays used ubiquitously in cQED systems for quantum non-demolition (QND) measurements of the qubit or resonator state\cite{Walter2017, Rosenblum2018}. In particular, it allows for QND measurements of the photon number parity of a resonator without revealing the underlying photon number distribution. This is, for example, a crucial error syndrome measurement in many error-correction schemes for quantum information encoded in harmonic oscillators\cite{Ofek2016, Hu2019}. While the basic physics of cQAD is in principle analogous to cQED, it has thus far been challenging for cQAD systems to reach a regime where the dispersive interaction is sufficiently large compared to decoherence to allow for such operations. This is largely because it is difficult to maintain state-of-the art coherence times for both qubits and mechanical resonators while combining them into a single device. Nevertheless, previous works have shown that cQAD systems based on bulk acoustic wave resonators (HBAR) are promising for achieving a good balance between quantum coherence and electromechanical coupling strength\cite{Chu2020}.

\begin{figure}
\centering
\includegraphics[width=8cm]{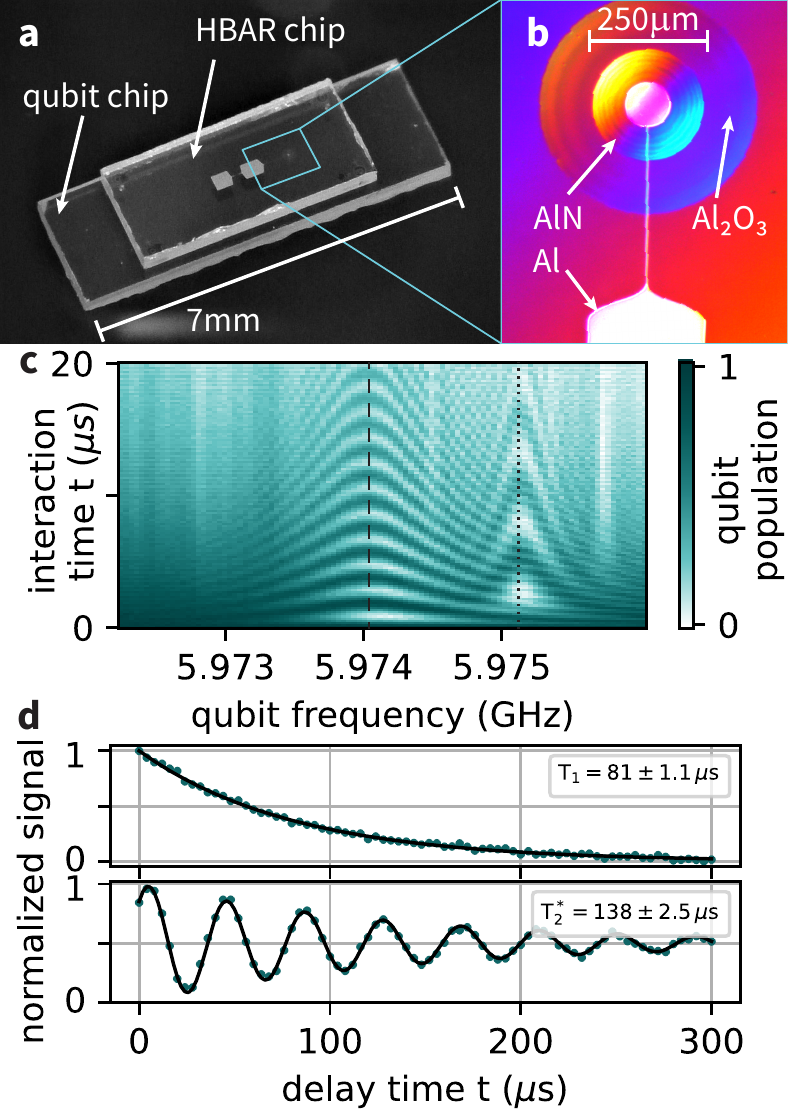}
\caption{\textbf{$\hbar$BAR device characterization.} 
\textbf{(a)} Photograph of the flip-chip bonded $\hbar$BAR device. \textbf{(b)} Optical microscope image of qubit antenna and HBAR. The circles around the qubit antenna are the edges of the dome feature etched into the substrate. The Newton rings visible on the convex acoustic resonator surface are used in the alignment process. The color is a result of the microscope light interfering between the chip surfaces. \textbf{(c)} Vacuum Rabi oscillations between qubit and phonon modes. The lines indicate the frequencies of the LG-00 (dashed) and LG-10 (dotted) modes.  \textbf{(d)} Energy relaxation (top) and Ramsey coherence measurement (bottom) of the LG-00 phonon mode. Green dots are datapoints, black solid lines are fits to exponential and exponential sine functions, respectively. 
The phase at the first point of the $T_2^*$ measurement is a result of the qubit-phonon detuning during the delay time combined with the finite duration of the SWAP operation. \label{fig1}}
\end{figure}

In this work, we experimentally demonstrate a cQAD system operating in the strong dispersive regime and use it to perform measurements of the phonon number distribution and parity of quantum states of a HBAR. 
Our system is a $\hbar$BAR-type device consisting of a plano-convex HBAR fabricated using piezoelectric aluminum nitride on a sapphire substrate and a superconducting qubit on a separate sapphire chip\cite{Chu2018}. 
We use a single-junction, non-flux tunable transmon qubit and house the $\hbar$BAR inside a superconducting Al cavity, which improves the coherence time of the qubit compared to previous devices. 
This improvement is crucial for reaching the strong dispersive regime. 
All frequency tuning is performed by applying microwave drives to Stark shift the qubit instead of applying an external magnetic field\cite{ChuScience2017}. The assembled $\hbar$BAR is shown in Figure \ref{fig1}a, with the 3D transmon visible through the HBAR chip. The two components are aligned and assembled using an industrial flip-chip bonder, which improves the reproducibility and robustness compared to devices used in previous works\cite{Satzinger2019}.
The magnified view in Figure \ref{fig1}b indicates good alignment between the qubit electrode and the HBAR.

The HBAR supports a complex and dense mode structure. 
The acoustic velocities and dimensions of the resonator result in a longitudinal free spectral range (FSR) of $\sim\unit{12}{MHz}$, while the frequency spacing between modes with different Laguerre-Gaussian (LG) transverse mode numbers is $\sim\unit{1}{MHz}$.
This puts constraints on the range of device parameters that would allow us to use the qubit for dispersive measurements of a single acoustic mode. 
First, we would like the coupling strength $g$ of the qubit to the LG-00 mode to be larger than to those with higher-order transverse mode numbers. 
Second, $g$ and the detuning $\Delta = \omega_q-\omega_m$ between the qubit frequency $\omega_q$ and phonon frequency $\omega_m$ should be much smaller than the FSR. 
If these constraints are satisfied, the effective Hamiltonian of our system in the dispersive regime of $\Delta \gg g$ can be approximated as\cite{blais2004cavity} 
\begin{linenomath*}
\begin{equation}
    H/\hbar \approx \omega_m a^{\dagger}a + \frac{1}{2}\left(\omega_q + \chi a^{\dagger}a\right) \sigma_z \;.
    \label{eq:dispHam}
\end{equation}
\end{linenomath*}
Here $a$ is the lowering operator for the acoustic mode, $\sigma_z$ is the Pauli z operator for the qubit, and $\frac{1}{2}\chi a^{\dagger}a\sigma_z$ is the dispersive interaction term that shifts the qubit frequency by $\chi a^{\dagger}a$, where \cite{Koch2007}
\begin{linenomath*}
\begin{equation}
   \chi = -2\frac{\abs{g}^2}{\Delta}\frac{\alpha}{\Delta-\alpha} \approx 2 \frac{\abs{g}^2}{\Delta} 
\label{eq:chi}
\end{equation}
\end{linenomath*}
is the qubit frequency shift per phonon.
The approximation is valid in our case, where $\Delta$ is much smaller than $\alpha$, the anharmonicity of the qubit.
We note that this is the opposite limit from the usual case in cQED, where typically $\Delta\gg\alpha$. 
Furthermore, to satisfy the requirement of $g\ll $ FSR, $g$ in our system is limited to $\lesssim\unit{1}{MHz}$, which is at least one order-of-magnitude smaller than in most cQED systems. Despite these constraints, we now show that the design and fabrication of our system leads to quantum coherences that are sufficient for operation in the strong dispersive regime.

To that end, we first characterize our device by measuring the coherence properties of the qubit and the phonon mode, as well as the coupling strength between them.  
Using standard cQED techniques\cite{krantz2019quantum, Blais2021cqed} we find for the qubit an energy relaxation time of $T_1^q=\unit{10.2\pm0.4}{\mu s}$, a Ramsey decoherence time of $T_2^{q*}=\unit{10.5\pm0.5}{\mu s}$, and an echo decoherence time of $T_{2e}^{q}=\unit{11.6\pm0.4}{\mu s}$.
Direct driving and readout of the qubit is performed at a detuning of $\Delta_{\mr{rest}}/2\pi=-\unit{4.1}{MHz}$ from the LG-00 phonon mode of interest. 
Tuning the qubit into resonance with the phonon mode then allows the two systems to coherently exchange energy through the resonant Jaynes-Cummings (JC) interaction, which is used to implement qubit-phonon SWAP operations for characterizing the phonon mode and initializing it in a quantum state in subsequent experiments\cite{ChuScience2017,Chu2018}.
The phonon mode spectrum and the coupling to the qubit are measured by exciting the qubit at $\Delta_{\mr{rest}}$, then tuning it to different frequencies and reading out the qubit state after a variable time (Figure \ref{fig1}c). 
We observe two oscillatory features centered around \unit{5.9741}{GHz} and \unit{5.9752}{GHz}, corresponding to vacuum Rabi oscillations between the qubit and the LG-00 and LG-10 phonon modes, respectively.
From these vacuum Rabi oscillations, we extract a coupling strength for the LG-00 mode of $g=\unit{259.5\pm0.3}{kHz}$. 

Using protocols described in previous work\cite{ChuScience2017}, we find $T_1=\unit{81\pm 1.1}{\mu s}$ and  $T_2^*=\unit{138\pm 2.5}{\mu s}$ for the phonon mode (Figure \ref{fig1}d), corresponding to a dephasing time of $T_{\phi} = \unit{932\pm 112}{\mu s}$. 
During the delay times for both measurements, we tune the qubit back to $\Delta_{\mr{rest}}$. 
This allows us to measure the intrinsic phonon coherence without a significant effect due to Purcell decay through the qubit. 
Note that both the qubit and phonon coherence times have increased significantly from previous devices\cite{Chu2018}.

\begin{figure}
\centering
\includegraphics[width=8cm]{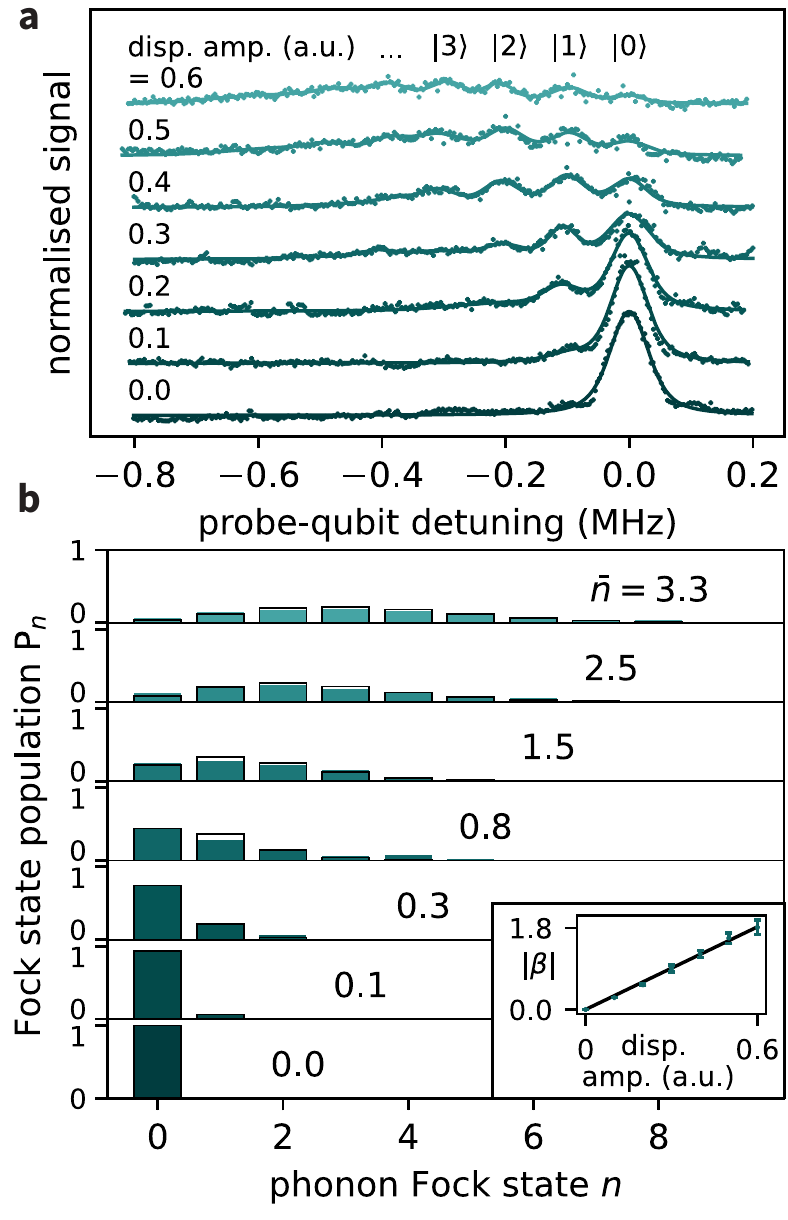}
\caption{\textbf{Dispersive measurement of phonon coherent states.}
\textbf{(a)} Spectroscopy of the qubit while dispersively coupled to coherent phonon states prepared using displacement drives with the indicated amplitudes. Solid lines are fits to sums of Voigt profiles, used to extract the Fock state populations $P_n$.
The data is rescaled so that the total population is normalized, and each trace is shifted vertically by 0.5 for clarity.
The \ket{0} peak is aligned for all spectra to correct for slow qubit frequency fluctuations.
\textbf{(b)} $P_n$ extracted from the measurements in panel \textbf{a} and corresponding mean phonon number $\bar{n}$ from fit to Poisson distribution.
Black boxes are the ideal Poisson distributions assuming an average phonon number of $\bar{n}$. 
Inset shows linear relation between the displacement drive amplitude and $|\beta|=\sqrt{\bar{n}}$. Error bars represent one standard deviation of the fit uncertainty of the Poisson distribution. 
\label{fig2} }
\end{figure}

The improved coherence of our system allows us to operate it in the strong dispersive regime, which we now demonstrate through phonon-number resolving measurements of the qubit spectrum.
We create coherent states in our phonon mode by driving it on resonance with a microwave pulse while the qubit is at $\Delta_{\mr{rest}}$.
After the phonon state preparation, we decrease the detuning to $\Delta_{\mr{coherent}}/2\pi=-\unit{1.2}{MHz}$ to reach a regime where the dispersive shift of the qubit frequency due to a single phonon should be much larger than the qubit's intrinsic linewidth $\gamma^{*}_2/2\pi=\unit{15.1}{kHz}$. 
We then perform qubit spectroscopy using a probe pulse with a bandwidth of \unit{10.6}{kHz}, less than $\gamma^*_2$, thus ensuring that the qubit linewidth is not significantly broadened. 
Finally, the qubit is read out at the frequency $\Delta_{\mr{rest}}$.
The resulting qubit spectra for different amplitudes of the displacement drive on the phonon modes is shown in Figure \ref{fig2}a. 
At larger amplitudes, we observe multiple resonances, as expected from the dispersive interaction with a superposition of phonon Fock states.
We fit each spectrum to a sum of Voigt profiles in order to extract the height of the peaks. 
These heights then give the population $P_n$ of each Fock state $\ket{n}$ after normalizing such that $\sum_n P_n=1$. 
From fitting a Poisson distribution to the measured populations, we obtain the mean phonon number $\bar{n}$ of the coherent state that most closely matches the data (see Figure \ref{fig2}b).
We find a linear dependence of the extracted coherent state amplitude $\abs{\beta} = \sqrt{\bar{n}}$ on the drive amplitude (see inset \ref{fig2}b), indicating the capability of performing coherent displacements of the phonon mode up to $\abs{\beta} = 1.9$. Additionally, these measurements provide a calibration between the drive amplitude we set and the resulting displacement amplitude (see Supplementary Information). Both of these results will be used in the experiments that follow.  

\begin{figure}
\centering
\includegraphics[width=7cm]{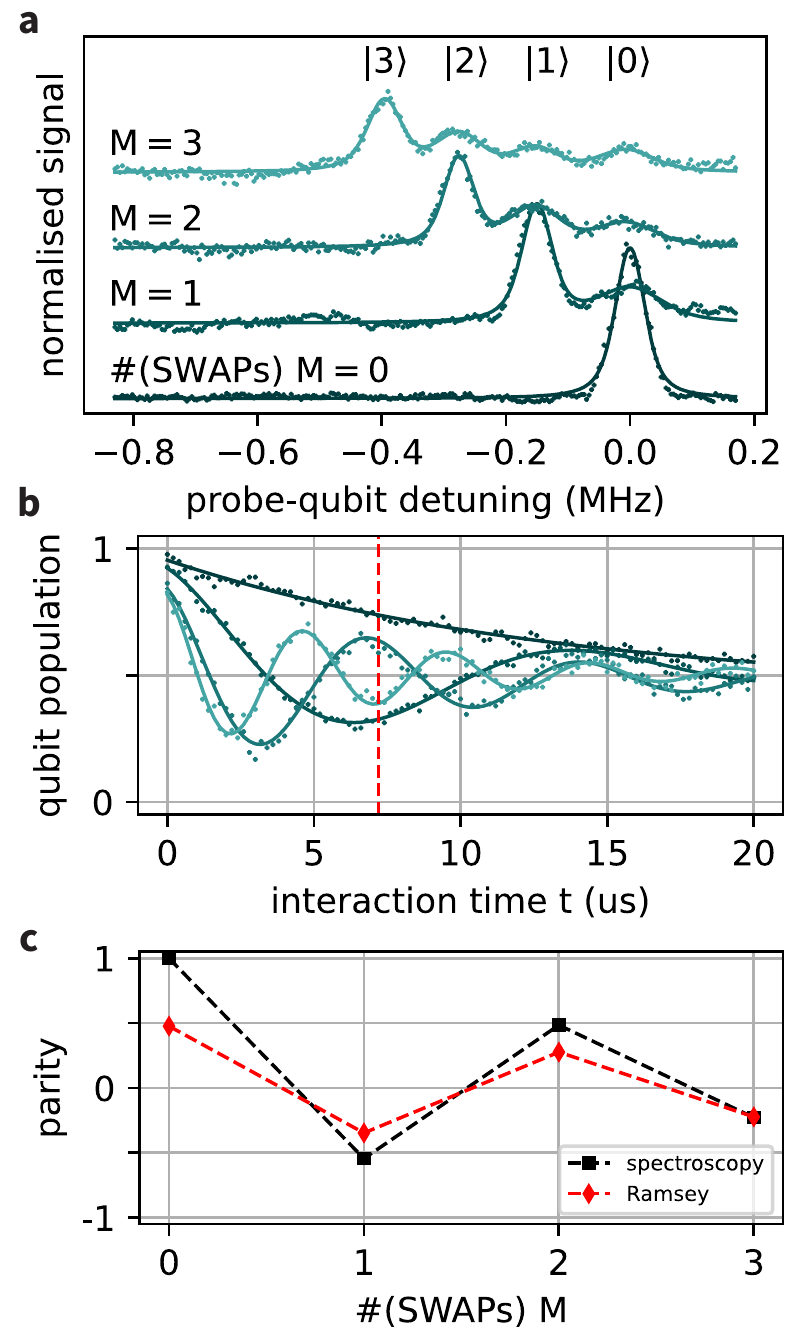}
\caption{
\textbf{Dispersive measurement of phonon Fock states.}
Spectroscopy \textbf{(a)} and Ramsey measurement \textbf{(b)} of the qubit while dispersively coupled to the phonon mode, following state preparation that ideally results in Fock states containing $M$ phonons. 
For \textbf{a}, the same procedure as in Figure \ref{fig2}a was used for fitting and rescaling the data to extract the phonon populations. 
In \textbf{b}, solid lines are fits with an exponentially decaying cosine. 
Oscillations do not start from unity due to a finite qubit population after the phonon state preparation.
The dashed red line indicates the Ramsey interrogation time used for measuring the parity of the prepared state.
\textbf{(c)} Parity of the prepared phonon Fock states measured by spectroscopy (black) and Ramsey (red) measurements. 
Dashed lines are guides to the eye. Error bars are smaller than markers. \label{fig3} }
\end{figure}

We now extend our analysis further to non-classical states prepared in the phonon mode. 
We repeat the measurement procedure adopted for Figure \ref{fig2}a, but now with the phonon mode initialized in Fock states containing $M$ phonons. 
We perform this initialization up to $M=3$ using repeated qubit excitations and SWAP operations with the phonon mode\cite{Chu2018}. 
In this case, we set the detuning during the dispersive interaction to be $\Delta_{\mr{Fock}}/2\pi=-\unit{0.8}{MHz}$. 
The smaller detuning leads to larger $\chi$ and better resolved peaks in the qubit spectrum at the expense of a less ideal dispersive approximation.
Figure \ref{fig3}a shows the qubit spectroscopy measurements, each resulting in one prominent peak corresponding to the target Fock state. 
Additionally, smaller peaks corresponding to finite population in lower Fock states can be seen, as expected from imperfect SWAP operations and phonon decay during preparation and measurement. 
Calculating the difference in frequency between the $\ket{0}$ peak for the $M = 0$ case and the $\ket{1}$ peak for the $M = 1$ case yields $\chi_{\mr{Fock}}/2\pi=-\unit{147}{kHz}$. 
We observe slightly smaller frequency differences between peaks corresponding to higher phonon populations, which we comment on in the Supplementary Information. 
The measured frequency differences are compatible with the dispersive shifts we expect from simulations. 
Again we confirm operation of our system in the strong dispersive regime.

The change in qubit frequency due to different phonon states can also be observed in the time domain by performing Ramsey-type measurements on the qubit\cite{Bertet2002, Sun2014}.
After initializing the phonon mode in a Fock state as above, we prepare the qubit in the superposition state $(\ket{g}+\ket{e})/\sqrt{2}$. 
We then move the qubit frequency from $\Delta_{\mr{rest}}$ to $\Delta_{\mr{Ramsey}}/2\pi=-\unit{1.9}{MHz}$ from the phonon mode, where we let it dispersively interact with the phonon state for a variable time $t$.
$\Delta_{\mr{Ramsey}}$ is chosen to be larger than $\Delta_{\mr{Fock}}$ to minimize deviations from the ideal dispersive Hamiltonian while still maintaining a large enough $\chi$.
Finally, we move the qubit back to $\Delta_{\mr{rest}}$, perform the second $\pi/2$-pulse, and read-out the qubit state. 
The phase of the second $\pi/2$-pulse is calibrated such that there is no oscillation in the Ramsey measurement corresponding to the phonon $\ket{0}$ state.
The measured data is shown in Figure \ref{fig3}b.  
We observe decaying oscillations at frequencies equal to the dispersive shift $M|\chi_{\mr{Ramsey}}|/2\pi=M\times \unit{70}{kHz}$. 
To explain this observation, we note that, in the frame rotating at the bare qubit frequency $\omega_q$, the initial superposition state evolves due to the dispersive interaction term into $(\ket{g}+\exp(-iM\chi_{\mr{Ramsey}} t)\ket{e})/\sqrt{2}$. 
After the second $\pi/2$ pulse of the Ramsey sequence, this relative phase is mapped onto the qubit population. 
The striking feature of this protocol is that for an interaction time $t_0=\pi/|\chi_{\mr{Ramsey}}|$, the measurement result in the ideal case is $\langle\sigma_z\rangle=\cos(M\pi)$, which yields the phonon number parity $\Pi$. 
In Figure~\ref{fig3}b, we can indeed observe that at time $t_0=\unit{7.1}{\mu s}$, an even (odd) Fock state results in a high (low) qubit population. We emphasize here that determining the parity with this procedure requires only a single measurement with interaction time $t_0$ and does not reveal the underlying phonon distribution. It is therefore more efficient and more useful for bosonic error correction protocols than the alternative method of computing the parity using the $P_n$ extracted from the full qubit spectra as $\Pi = \sum_n (-1)^n P_n$. We compare the two methods in Figure \ref{fig3}c. As expected, the prepared even (odd) Fock states show positive (negative) parity for both the spectroscopy and the Ramsey-type measurement. For higher phonon Fock states, both methods show parities close to zero, which we attribute to imperfect state preparation and decoherence, as explained earlier. We also note here that a non-negligible $g/\abs{\Delta_{\mr{Ramsey}}} \approx 0.14$ limits the QND nature of our current measurements. Improving this figure of merit will be the subject of our future work.

\begin{figure}
\centering
\includegraphics[width=16cm]{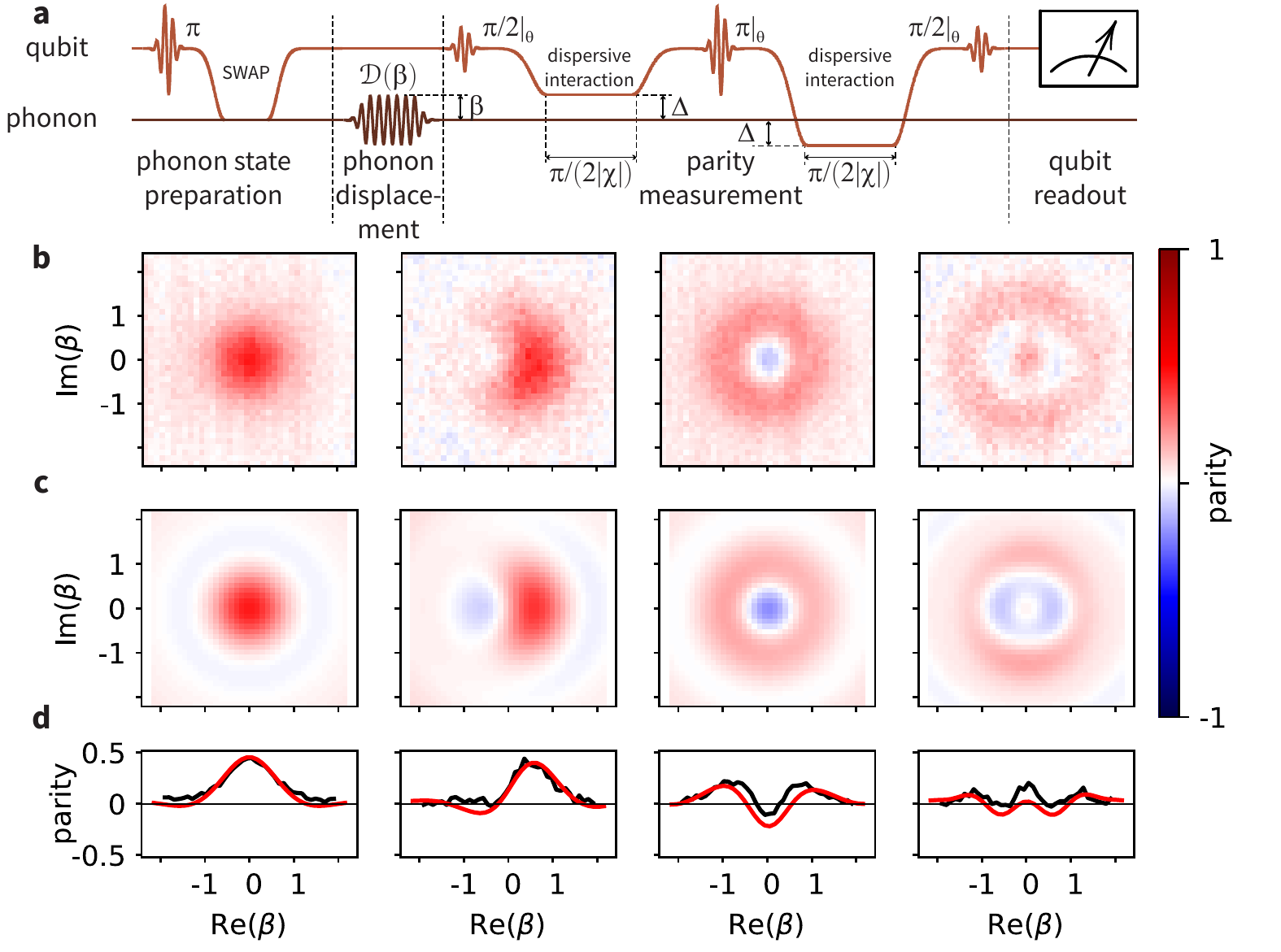}
\caption{
\textbf{Wigner tomography of non-classical phonon states.}
\textbf{(a)} Sequence used for Wigner tomography of a phonon mode prepared in \ket{1}. See main text for details.
\textbf{(b)} Measured Wigner functions of the phonon mode prepared in, from left to right, $\ket{0}$, $\frac{1}{\sqrt{2}}(\ket{0}+\ket{1})$, $\ket{1}$, and $\ket{2}$. 
Axes are calibrated using the data in Figure \ref{fig2} (see Supplementary information). 
\textbf{(c)} Wigner functions obtained from master equation simulations of the JC Hamiltonian. The axes are calibrated by fitting the simulated density matrix of the phonon mode after a displacement.
\textbf{(d)} Cuts along the Im($\beta$) = 0 axis for measured results shown in panel \textbf{b} (black) and simulation results shown in panel \textbf{c} (red). \label{fig4}}
\end{figure}

The measurement of  phonon state parity together with coherent displacements gives us the necessary tools to perform Wigner tomography. 
This is enabled by the fact that the Wigner function can be expressed as displaced parity measurements\cite{RoyerPRA1977}. 
We again emphasize that, in contrast to previous work where the parity was determined by first extracting the phonon number distribution from time-resolved measurements of the resonant qubit-phonon interaction\cite{Chu2018}, here we are able to measure the parity at each position in phase space with a single averaged measurement on the qubit.
Figure \ref{fig4}a shows the sequence used to perform Wigner tomography of the phonon mode in the \ket{1} state (for other states, the sequence differs only in the state preparation step at the beginning). This is followed by a coherent displacement of the phonon state by the complex amplitude $\beta$.
Our device suffers from slow qubit frequency fluctuations on the $\sim \unit{10}{kHz}$ frequency scale common in superconducting qubits\cite{burnett2019decoherence}. 
In contrast to most cQED systems, however, this is comparable to our dispersive shift.
Therefore, we improve our parity measurement protocol by introducing a dynamical decoupling pulse sequence to mitigate the effect of low frequency noise as follows:
We initialize the qubit in the state $(\ket{g}+\ket{e})/\sqrt{2}$ and then tune its frequency to $\Delta_{\mr{Ramsey}}$. 
After an interaction time of $\pi/(2|\chi_{\mr{Ramsey}}|)= \unit{3.53}{\mu s}$, the qubit superposition has acquired a relative phase of $n \pi / 2$. 
We then move the qubit frequency back to its rest point, apply a $\pi$-pulse, and move the qubit frequency to a detuning of $-\Delta_{\mr{Ramsey}}$, flipping the sign of the dispersive shift. 
In combination with the echo pulse, this results in a total relative phase accumulation of $n \pi$ after another interaction time $\pi/(2|\chi_{\mr{Ramsey}}|)$.
The relative phase that changes due to low frequency noise (such as shot-to-shot qubit frequency jumps) does not depend on the sign of the detuning and is thus canceled out.
The echo sequence ends by taking the qubit frequency back to $\Delta_{\mr{rest}}$ and applying a second $\pi/2$-pulse, whose phase is calibrated in order to compensate the effect caused by the Stark shift drive used to change the qubit frequency.
We also take additional measures to mitigate the effects of finite $g/\Delta$ on our Wigner tomography. The first order effect is a $\beta$-dependent deviation from an ideal parity measurement and is canceled by averaging data taken with four different qubit drive phases, indicated by $\theta$ in Figure~\ref{fig4}a. The remaining effect is an interaction-time dependent constant offset of the Wigner function that oscillates with a frequency much faster than $\abs{\chi}$. This allows us to precisely choose the interaction time in order to minimize this effect. For details see the Supplementary Information.

Figure \ref{fig4}b shows measured Wigner functions for the phonon mode prepared in the states $\ket{0}$, $(\ket{0}+\ket{1})/\sqrt{2}$, $\ket{1}$, and $\ket{2}$. 
For comparison, Figure \ref{fig4}c shows Wigner functions obtained from a master equation simulation\cite{johansson2012qutip} of our experimental sequence using the full JC Hamiltonian, including finite pulse lengths and both decay and decoherence in the system.
The results are in good agreement with the measurements and reproduce features that would not be present in the ideal case. 
For example, note that the Wigner function for $\ket{2}$ does not show circular symmetry due to imperfect state preparation, and the non-zero values of the Wigner function at large displacements can again be attributed to a finite $g/\abs{\Delta}$.
The discrepancy of both the measurements and simulations from the ideal Wigner functions, however, is mainly due to qubit and phonon decay and decoherence.

Our results show that dispersive measurements of mechanical quantum states are now possible in $\hbar$BAR-type cQAD systems. We achieved this through improvements to the coherence properties of both the qubit and the phonon mode in order to reach the strong dispersive regime. We carefully chose the qubit-phonon detuning in order to reach a compromise between ensuring that the dispersive interaction is strong enough to allow for coherent operations and remaining in the dispersive regime. For the purpose of performing quantum state tomography, we devised protocols that mitigated to first order the effect of qubit-phonon energy exchange due to finite $g/\Delta$ and qubit frequency fluctuations that are non-negligible compared to $\chi$. However, the decoherence and energy exchange during the measurement will still degrade the quality of single-shot, QND measurements that are needed in, for example, error correction protocols. 
Therefore, continued improvements to the basic device properties of $\hbar$BAR devices remain crucial, and further exploration of the large space of design parameters and their effect on device performance is the subject of our ongoing work (see Supplementary Information). 

While quantum operations that make use of the dispersive interaction are useful ingredients for quantum information processing, they can be also seen as the first steps toward a much broader range of future experiments and applications, some of which are unique to mechanical systems. Quantum states of mechanical resonators such as Fock state superpositions and Schrödinger cat states have been proposed as resources for quantum-enhanced frequency or force sensing\cite{McCormick2019} and testing modifications to quantum mechanics at macroscopic scales\cite{Penrose1996, Gely2021}. Many existing protocols for preparing such complex quantum states make direct use of the dispersive interaction\cite{Vlastakis2013, Heeres2017}. At the same time, other protocols for quantum control of bosonic modes, such as parametrically driven multimode gates\cite{Gao2019, Hann2019} or autonomous stabilization of nontrivial quantum states\cite{Leghtas2015}, do not directly make use of the dispersive Hamiltonian. However, the requirements on coupling strengths and coherences in order to perform these operations are effectively similar to the strong dispersive regime. Our results suggest that these demonstrations will soon be within reach for cQAD devices.

\bibliography{disp_parity}

\section*{Acknowledgements}

We thank Xiaobao Cao and Andrew deMello for help with the flip-chip bonding process and Jean-Claude Besse for help with qubit fabrication. We thank Boxi Li for providing support with QuTiP simulations. Fabrication of devices was performed at the FIRST cleanroom of ETH Zürich and the BRNC cleanroom of IBM Zürich.

\section*{Author contributions statement}

U.v.L. and L.M. designed and fabricated the device. U.v.L. and M.B. made the parametric amplifier used for qubit readout. M.B. wrote experiment control software. Y.Y., U.v.L, and M.B. performed the experiment and analyzed the data. Y.Y. performed QuTiP simulations of the experiment. M.F., M.B., and Y.C. performed theoretical calculations. Y.C. supervised the work. U.v.L., Y.Y., M.B., M.F., and Y.C. wrote the manuscript.

\section*{Additional information}

\textbf{Competing interests} The authors declare no competing interests. \\
\textbf{Correspondence and requests for materials} should be addressed to U.v.L. or Y.C.

\clearpage
\newpage

\renewcommand{\thesection}{}  
\renewcommand{\thesubsection}{\Alph{subsection}}  

\renewcommand{\thetable}{S\arabic{table}}  
\renewcommand{\thepage}{S\arabic{page}}  
\renewcommand{\thefigure}{S\arabic{figure}}
\renewcommand{\theequation}{S\arabic{equation}}
\setcounter{page}{1}
\setcounter{figure}{0}
\setcounter{table}{0}
\setcounter{section}{0}
\setcounter{equation}{0}

\resetlinenumber

\begin{center}
    \section*{Supplementary information for \\ Parity measurement in the strong dispersive regime of circuit quantum acoustodynamics}
\end{center}
    Uwe von Lüpke$^{1,2,\dagger,*}$, Yu Yang$^{1,2,\dagger}$, Marius Bild$^{1,2,\dagger}$, Laurent Michaud$^{1,2}$, Matteo Fadel$^{1,2}$, Yiwen Chu$^{1,2,**}$\\
    $^1$ \textit{Department of Physics, ETH Zürich, 8093 Zürich, Switzerland} \\
    $^2$ \textit{Quantum Center, ETH Zürich, 8093 Zürich, Switzerland} \\
    $^\dagger$ these authors contributed equally to this work\\
    $^*$ vluepkeu@phys.ethz.ch\\
    $^{**}$ yiwen.chu@phys.ethz.ch
\tableofcontents
\subsection{Sample design and fabrication} 
The transmon qubit and the HBAR are fabricated on two separate chips. 
The fabrication procedure for each element is identical to that in previous work\cite{Chu2018}. 
However, the combination of the two chips into a flip-chip device follows a new method. 
We lithographically define spacers made of the epoxy-based photoresist SU-8 on the HBAR chip, similar to what has been demonstrated in Ref.~\cite{conner2021superconducting}. 
Before combining the two chips using a commercial bonder (SET - FC150), we apply liquid adhesive to the surface of the qubit chip using a PDMS stamp. Both the spacers and adhesive are close to the corners of the chip and far away from the device region to avoid additional dielectric loss for the qubit.
We align the two chips by matching the qubit electrode to the Newton rings that are visible in the aluminum nitride dome (see Figure~\ref{fig1}b).  
More details of the fabrication process will be described in a future publication. 

In the process of fabricating and characterizing devices for this work, we have gained a better understanding of how various parameters of the system are affected by materials and design choices. The flip-chip geometry allows us to systematically compare the coherences of our standard transmon qubits, which typically have $T_1$'s of 30-\unit{50}{\mu s}, to those that have been bonded to HBARs, which typically have $T_1 < \unit{20}{\mu s}$. The decrease in qubit $T_1$ after incorporation into a $\hbar$BAR is consistent with the results from earlier works\cite{Chu2018}, despite the use of different materials and techniques for flip-chip bonding. This strongly suggests that the piezoelectric material introduces a dominant loss channel for the qubit, likely through a combination of local dielectric loss in the material and piezoelectric coupling to unconfined phonon modes\cite{Scigliuzzo2020}. We have also observed the coupling strengths of the qubit to the different transverse modes can be modified by changing the piezoelectric material, qubit geometry, spacing and alignment between the two bonded chips, etc. 

\newpage
\subsection{Device and experimental parameters.}
Table \ref{tab:device} gives the device parameters for the transmon and the HBAR resonator. 

\begin{table}[H]
\centering
\begin{tabular}{lr|lr}
\hline
parameter & value & parameter & value \\
\hline
$\omega_q/2\pi$ &  5.9762\,GHz &$\omega_m^{\mr{LG-00}}$ & 5.9741\,GHz \\
$E_C/h$ &  214\,MHz &$\omega_m^{\mr{LG-10}}$ & 5.9752\,GHz  \\
$E_J/h$ &  22.4\,GHz &HBAR FSR & 12\,MHz \\
$\gamma_1/2\pi$ @ $\Delta_{\mr{rest}}$ & 15.6\,kHz &$\kappa_1/2\pi$  @  $\Delta_{\mr{rest}}$ & 2.0\,kHz \\
$\gamma_2^{*}/2\pi$ @ $\Delta_{\mr{rest}}$ &15.1\,kHz &$\kappa_2^*/2\pi$ @ $\Delta_{\mr{rest}}$ & 1.2\,kHz \\
$\gamma_2^{E}/2\pi$ @  $\Delta_{\mr{rest}}$ & 13.7\,kHz &$\kappa_1/2\pi$ @ $\Delta_{\mr{Ramsey}}$ & 2.6\,kHz \\
$\gamma_1/2\pi$ @  $\Delta_{\mr{Ramsey}}$ & 12.1\,kHz &$\kappa_2^*/2\pi$ @ $\Delta_{\mr{Ramsey}}$ & 2.1\,kHz \\
$\gamma_2^{*}/2\pi$ @  $\Delta_{\mr{Ramsey}}$ &15.7\,kHz &$\Delta_{\mr{rest}}/2\pi$ &  -4.1\,MHz \\
$\gamma_2^{E}/2\pi$ @ $\Delta_{\mr{Ramsey}}$ & 12.7\,kHz &$\Delta_{\mr{coherent}}/2\pi$ &  -1.2\,MHz \\
$g^{\mr{LG-00}}/2\pi$ & 259.5\,kHz &$\Delta_{\mr{Fock}}/2\pi$ &  -0.8\,MHz\\
$g^{\mr{LG-10}}/2\pi$& 91.3\,kHz & $\Delta_{\mr{Ramsey}}/2\pi$ &  -1.9\,MHz \\
\hline
\end{tabular}
\caption{\textbf{List of qubit and phonon properties and experimental parameters.} The qubit frequency $\omega_q$ is measured at zero Stark shift. For other parameters, we denote the detuning between qubit and LG-00 phonon mode at which they are measured.
$E_C$ is calculated from the qubit anharmonicity $\alpha$, which is measured spectroscopically by probing the two-photon transition between the qubit's \textit{g} and \textit{f} state. 
Then $E_C\approx \alpha=2E_{ge}-E_{gf}$. 
The energy relaxation and decoherence rates of the qubit $\gamma$ and of the LG-00 phonon mode $\kappa$ are measured in the time domain.}
\label{tab:device}
\end{table}
\newpage

\subsection{Coherent state amplitude calibration.}\label{supp:alpha_calib}
\begin{figure}
\centering
\includegraphics[width=16cm]{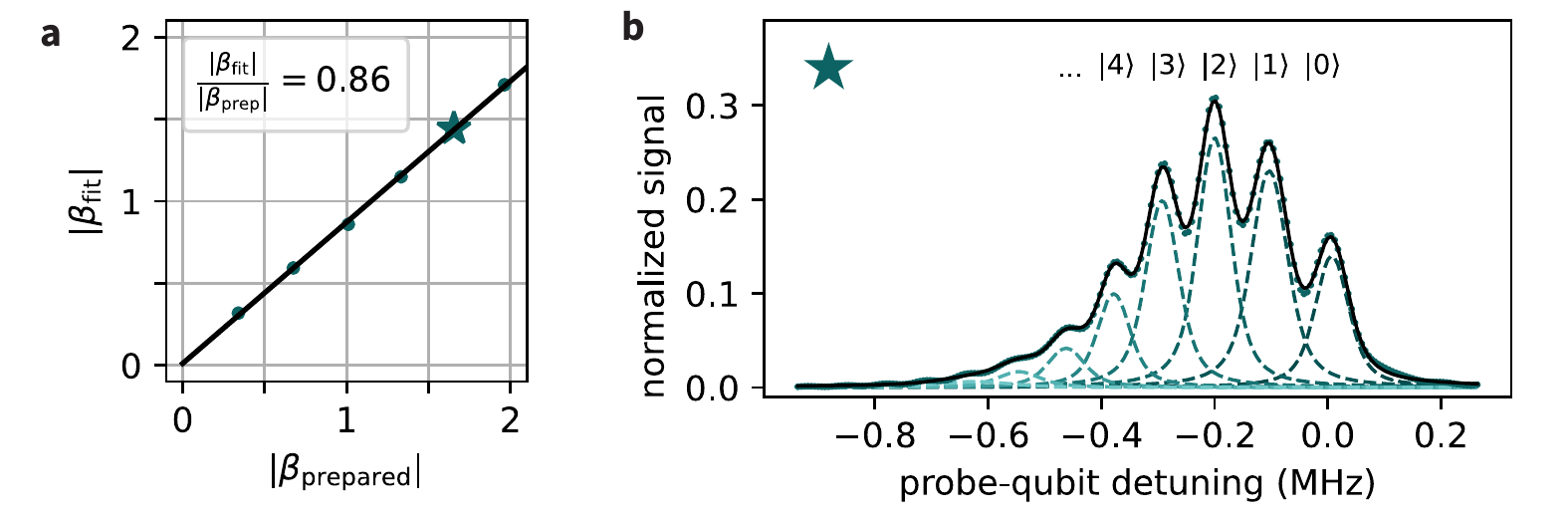}
\caption{
\textbf{Coherent state amplitude simulation.} 
\textbf{(a)} Linear relationship between the prepared coherent state amplitude $|\beta_\mr{prep}|$ and the amplitude $|\beta_\mr{fit}|$ extracted from the simulated spectroscopy. The star indicates the simulation for which the Voigt fit is shown in (b). 
\textbf{(b)} Example of a qubit spectroscopy simulation with fit to a sum of Voigt profiles. The relative height of each peak shows the population of the corresponding phonon Fock state. The data is rescaled to normalize the total population. Fitting the extracted populations to a Poisson distribution yields $|\beta_\mr{fit}|=1.44$. 
\label{figS:alpha_calib} }
\end{figure}
In Figure \ref{fig2} we show spectroscopic measurements of the qubit while it is dispersively coupled to coherent phonon states with different amplitudes $|\beta|$. 
The coherent states are prepared using a displacement drive on resonance with the phonon mode. As described in the main text, we use the linear relationship between the amplitude of the displacement drive and that of the measured coherent state to calibrate the axes in the Wigner tomography shown in Figure \ref{fig4}b. 

During the spectroscopic measurement of the qubit, however, the coherent state amplitude $|\beta|$ decays due to energy relaxation of the phonon, such that the measured coherent state amplitude $|\beta_\mr{fit}|$ is smaller than the prepared $|\beta_\mr{prep}|$. 
To quantify the effect of this decay, we perform time-domain simulations of the phonon displacement and qubit spectroscopy sequence. 
We then acquire $|\beta_\mr{prep}|$ from the density matrix of the prepared phonon state directly after the simulated displacement drive. 
To acquire $|\beta_\mr{fit}|$, we perform a fit of the simulated qubit spectropscopy to a sum of Voigt profiles to extract the Fock state populations, as we did for the experimental data. An example of this is shown in Figure \ref{figS:alpha_calib}b.
Fitting the populations to a Poisson distribution then yields $|\beta_\mr{fit}|$. 
Figure \ref{figS:alpha_calib}a shows the linear relationship between the prepared and measured coherent state amplitudes corresponding to a ratio of $|\beta_\mr{fit}|/|\beta_\mr{prep}|=0.86$.

We can compare this to an estimate of the expected decay of $\abs{\beta}$ during the qubit spectroscopy. Taking into account the Purcell decay of the phonon through the qubit, the phonon decay rate at $\Delta_\mr{coherent}$ is $\kappa_1^\mr{coherent}/2\pi\approx\unit{3.2}{kHz}$. We can then estimate the ratio of measured to prepared $\abs{\beta}$ for a $\tau_\mr{spec}=\unit{15}{\mu s}$ long spectroscopy pulse as 
\begin{linenomath*}
\begin{equation}
    \frac{1}{\tau_\mr{spec}}\int_0^{\tau_\mr{spec}} e^{-\kappa_1^\mr{coherent}t} dt = 0.86, \;
\end{equation}
\end{linenomath*}
which agrees well with the simulated value.
We use the ratio $|\beta_\mr{fit}|/|\beta_\mr{prep}|=0.86$ to correct the axes of the Wigner tomography plots.
\newpage

\subsection{Deviations from dispersive Hamiltonian and effect on Wigner functions.}

In this section, we illustrate how deviations from the dispersive Hamiltonian due to terms of higher order in $g/\Delta$ affect the parity measurement implementation.
In particular, these terms give rise to a nontrivial background in the measured Wigner function. 
We show that these effects can be partially mitigated by averaging over measurements taken with different phases for the qubit drive pulses (angle $\theta$ in Figure~\ref{fig4}) and choosing an appropriate dispersive interaction time.

We start with some relevant expressions that will be needed in the calculation we are going to present. First, we find a basis transformation which diagonalizes the full Hamiltonian describing our system. The latter reads
\begin{linenomath*}
\begin{equation}
    H = \oa \adg a + \oq\dfrac{\sz}{2} + \underbrace{ g\spl a + g^\ast\smn\adg}_{\equiv\HJC} \;, \label{eq:fullH}
\end{equation}
\end{linenomath*}
where $\HJC$ is the (non-diagonal) Jaynes–Cummings interaction term describing the coupling  between qubit and phonon with strengh $g$.
Hamiltonian~\ref{eq:fullH} can be diagonalized perturbatively to first order in the interaction $\HJC$ through the Schrieffer–Wolff (SW) transformation. To this end, we define the operator
\begin{linenomath*}
\begin{align}
    U &\equiv \exp\left[ \dfrac{g}{\D} \spl a - \dfrac{g^\ast}{\D} \smn \adg \right] = e^A \notag\\
    &\simeq \mathbb{I} + \e \spl a - \e^\ast \smn \adg - \dfrac{|\e|^2}{2}\left( \spl\smn a\adg + \smn\spl \adg a \right) + O(\e^3) \;, 
\end{align}
\end{linenomath*}
where $\D=\oq-\oa$ is the detuning between the qubit and the acoustic frequencies, and $\e\equiv g/\D$. The transformed Hamiltonian takes the form
\begin{linenomath*}
\begin{align}
    H^\prime &\equiv U H U^\dagger \notag\\
    &= H + [A,H] + \dfrac{1}{2} [A,[A,H]] + \dfrac{1}{6} [A,[A,[A,H]]] + ... \notag\\
    &= H + \oa \dfrac{\HJC}{\Delta} - \oq \dfrac{\HJC}{\Delta} + \dfrac{1}{2}\left( \omega_a \dfrac{H_2}{\Delta} - \oq \dfrac{H_2}{\Delta} + 2 H_2 \right) + O(\e^3) \notag\\
    &= \oa \adg a + \oq\dfrac{\sz}{2} + \dfrac{1}{2} H_2 + O(\e^3) \notag\\
    &= \oa \adg a + \oq\dfrac{\sz}{2} +  \dfrac{|g|^2}{\Delta}\dfrac{\sz}{2}\left( \id + 2 \adg a \right) + O(\e^3) \;. \label{eq:HSW}
\end{align}
\end{linenomath*}
where in writing the third line we defined $H_2 \equiv \dfrac{|g|^2}{\Delta}\sz\left( \id + 2 \adg a \right)$ and used the relations
\begin{linenomath*}
\begin{align}
    U \adg a U^\dagger &= \adg a + \dfrac{\HJC}{\Delta} + \dfrac{1}{2}\dfrac{H_2}{\Delta} + O(\e^3) \\
    U \sz U^\dagger &= \sz - 2 \dfrac{\HJC}{\Delta} - \dfrac{1}{2}\left(2\dfrac{H_2}{\Delta}\right) + O(\e^3)   \\
    U \HJC U^\dagger &= \HJC + H_2 + O(\e^3)  \;, 
\end{align}
\end{linenomath*}
that are derived from the commutators
\begin{linenomath*}
\begin{align}
    [A,\HJC] &= \dfrac{|g|^2}{\Delta} \left( \sz + 2\sz \adg a + 1 \right) \\
    [A,\sz \adg a] &= \dfrac{g}{\Delta}\left( -2\spl a\adg a +  \sz\spl a  \right) + \dfrac{g^\ast}{\Delta}\left( -2\smn \adg\adg a +  \sz\smn \adg  \right) \;.
\end{align}
\end{linenomath*}

Note that in Equation~\ref{eq:HSW}, the term $\HJC$ is missing, as expected from the fact that the SW transformation was chosen to result in a Hamiltonian that is diagonal at least to first order in $\epsilon$. In fact, here the Hamiltonian is also diagonal to second order, which gives the effective dispersive Hamiltonian in Equation 1 of the main text.

It is now convenient to further transform $H^\prime$ into a frame rotating at frequency $\oq^\prime=\oq+\frac{|g|^2}{\Delta}$, for both the phonon and qubit degrees of freedom. Note that this is the frame of the actual measured qubit frequency. Taking $R=e^{i\oq^\prime \sz t/2}e^{i\oq^\prime \adg a t}$, the resulting Hamiltonian reads
\begin{linenomath*}
\begin{align}
    H^\prime_R &\equiv R H^\prime R^\dagger + i \frac{\partial R}{\partial t} R^\dagger \notag\\
    &= H^\prime - \oq^\prime \dfrac{\sz}{2} - \oq^\prime \adg a \notag\\
    &= (\oa -\oq^\prime) \adg a + \dfrac{|g|^2}{\Delta} \sz \adg a \notag\\
    &= - \Delta^\prime \adg a + \dfrac{\chi}{2} \sz \adg a \;, \label{eq:HSWR}
\end{align}
\end{linenomath*}
where we introduced $\Delta^\prime \equiv \Delta + \dfrac{|g|^2}{\Delta}$, and $\chi \equiv 2 |g|^2 / \Delta$. Similarly, the SW transformation acts on the basis states as
\begin{linenomath*}
\begin{subequations}
\begin{align}
    U\ket{g,n} &= \ket{g,n} + \e \sqrt{n} \ket{e,n-1} - \dfrac{|\e|^2}{2} n \ket{g,n} + O(\epsilon^3) \label{eq:Ug} \\ 
    U\ket{e,n} &= \ket{e,n} - \e^\ast \sqrt{n+1} \ket{g,n+1} - \dfrac{|\e|^2}{2} (n+1) \ket{e,n} + O(\epsilon^3)\;\label{eq:Ue}.
\end{align}
\end{subequations}
\end{linenomath*}
Note that the usual dispersive approximation simply treats the dispersive shift as the lowest order perturabtive correction to the energy of the un-transformed eigenstates $\ket{g, n}$ and $\ket{e,n}$. Here we will instead also keep the corrections to the eigenstates that are first and second order in $\e$, as shown in Equations~\ref{eq:Ug} and \ref{eq:Ue}.

\textbf{Ramsey sequence:} To concisely illustrate the effect of the higher order terms in $\epsilon$, we now compute the results for the Ramsey-type measurement of the parity. This consists of a $\pi/2$ pulse to the qubit, followed by an evolution for time $t_0=\pi/\abs{\chi}$ with $\chi=2 \abs{g}^2/\Delta$, and then a second $\pi/2$ pulse. Finally, a measurement of $\langle \sz\rangle$ is performed.

We have also verified that the salient features remain the same for the echo-type measurements used in Figure \ref{fig4} of the main text, but do not present the full calculation here. 
We instead briefly describe it in the next subsection.

Initially, the qubit is in the $\ket{g}$ state, and the joint state of qubit-phonon system is
\begin{linenomath*}
\begin{equation}\label{eq:psi0Ramsey}
    \ket{\psi_0} = \sum_n c_n \ket{g,n} \;.
\end{equation}
\end{linenomath*}
We now define a $\pi/2|_{\theta}$ pulse on the qubit as a counterclockwise rotation $R=e^{-i \frac{\eta}{2}\Vec{u}.\Vec{\sigma}}$ by $\eta=\pi/2$ around the $\Vec{u}=-\cos(\theta)\Vec{e}_y-\sin(\theta)\Vec{e}_x$ direction on the Bloch sphere, taking for example $\ket{g} \rightarrow \ket{g}+e^{i\theta}\ket{e}$.
Therefore, after the first $\pi/2|_{\theta}$ pulse to the qubit, the state reads
\begin{linenomath*}
\begin{equation}
    \ket{\psi_\theta} = \dfrac{1}{\sqrt{2}} \sum_n c_n \left( \ket{g,n} + e^{i\theta} \ket{e,n} \right) \;.
\end{equation}
\end{linenomath*}
After the SW transformation, the qubit state becomes
\begin{linenomath*}
\begin{align}
    &\ket{\psi_\theta^\prime} = U \ket{\psi_\theta} \notag\\ 
    &= \dfrac{1}{\sqrt{2}} \sum_n c_n \left( \ket{g,n} + \epsilon \sqrt{n} \ket{e,n-1} + e^{i\theta} \left( \ket{e,n} - \epsilon^\ast \sqrt{n+1} \ket{g,n+1}  \right)\right) + O(\epsilon^2) \;.
\end{align}
\end{linenomath*}
For simplicity, in writing this and the following results we will omit terms of order higher than $\epsilon$ and present the result of order $\epsilon^2$ only at the end.
After evolving this state for time $t_0=\pi/\abs{\chi}$ according to the Hamiltonian $H^\prime_R$, the state reads
\begin{linenomath*}
\begin{align}
    \ket{\psi_\theta^\prime(t=t_0)} &= e^{-i H^\prime_R t_0} \ket{\psi_\theta^\prime} \notag\\ 
    &= \dfrac{1}{\sqrt{2}} \sum_n c_n \left[ e^{i n \pi (\Dpr/\abs{\chi}\pm1/2)}  \ket{g,n} + \epsilon \sqrt{n} e^{i (n-1) \pi (\Dpr/\abs{\chi}\mp1/2)} \ket{e,n-1} \right. \notag\\
    &\phantom{\simeq\dfrac{1}{\sqrt{2}}\sum_n AA} \left. + e^{i\theta} \left( e^{i n \pi (\Dpr/\abs{\chi}\mp1/2)}\ket{e,n} - \epsilon^\ast \sqrt{n+1} e^{i (n+1) \pi (\Dpr/\abs{\chi}\pm1/2)} \ket{g,n+1}  \right)\right] \notag \\
    &\phantom{\simeq\dfrac{1}{\sqrt{2}}\sum_n}+O(\epsilon^2)\\
    &= \dfrac{1}{\sqrt{2}} \sum_n c_n e^{i n (\phi\pm\pi/2)} \left[\vphantom{e^{i n \pi (\Dpr/\abs{\chi}\pm1/2)}}\ket{g,n} + \epsilon \sqrt{n} e^{-i n \pi} e^{-i (\phi \mp\pi/2)} \ket{e,n-1} \right. \notag \\
    &\phantom{\simeq\dfrac{1}{\sqrt{2}}\sum_n} \left. + e^{i\theta} \left( e^{-i n \pi} \ket{e,n} - \epsilon^\ast \sqrt{n+1} e^{i (\phi\pm\pi/2)} \ket{g,n+1}  \right)\right] +O(\epsilon^2)  \label{eq:spaghetti}
\end{align}
\end{linenomath*}
where in the last line we defined $\phi(t) \equiv \Dpr t$, so $\phi(t_0) = \pi \Dpr/\abs{\chi}$. For simplicity we only consider $t = t_0$ here and write $\phi(t_0)$ as $\phi$, but it should be understood that the results are dependent on $t$, which will become important in Section \ref{supp:t}. The top (bottom) signs are for $\chi > 0\,~(\chi<0)$.
Undoing the SW transformation gives us the state expressed in the original basis of Equation~\ref{eq:psi0Ramsey}:
\begin{linenomath*}
\begin{align}
    \ket{\psi_\theta(t=t_0)} &= U^\dagger \ket{\psi^\prime_\theta(t=t_0)} \notag\\ 
    &= \dfrac{1}{\sqrt{2}} \sum_n c_n e^{i n (\phi\pm\pi/2)} \left[\vphantom{e^{(\Dpr/\abs{\chi}\pm1/2)}} \ket{g,n} + \e\sqrt{n} \left( e^{-i n \pi} e^{-i (\phi \mp\pi/2)} - 1 \right) \ket{e,n-1} \right. \notag \\
    &\phantom{\simeq\dfrac{1}{\sqrt{2}}\sum_n} \left. + e^{i\theta} \left( e^{-i n \pi} \ket{e,n} - \e^\ast \sqrt{n+1} \left(e^{i (\phi\pm\pi/2)} -e^{-i n \pi} \right) \ket{g,n+1}  \right)\right] + O(\epsilon^2) \;.
\end{align}
\end{linenomath*}
After the second $\pi/2|_\theta$ that concludes the Ramsey sequence, the state is
\begin{linenomath*}
\begin{align}
    \ket{\psi_f} &= \dfrac{1}{2} \sum_n c_n e^{i n (\phi\pm\pi/2)} \left[\vphantom{e^{ (\Dpr/\abs{\chi}\pm1/2)}} \left(1-e^{-i n \pi}\right)\ket{g,n} + e^{i\theta}\left(1+e^{-i n \pi}\right) \ket{e,n}\right. \notag\\
    &\phantom{=\dfrac{1}{2} \sum_n}+ \e\sqrt{n} \left( e^{-i n \pi} e^{-i (\phi \mp\pi/2)} - 1 \right) (\ket{e,n-1}-e^{-i\theta}\ket{g,n-1})  \notag \\
    &\phantom{=\dfrac{1}{2} \sum_n} \left. + e^{i\theta} e^{-i n \pi} \left(\e^\ast \sqrt{n+1} \left( 1 - e^{i n \pi} e^{i (\phi\pm\pi/2)} \right) (\ket{g,n+1}+e^{i \theta}\ket{e,n+1})  \right)\right] + O(\epsilon^2) \;.
\end{align}
\end{linenomath*}
For later convenience, we express this state as two sums over even and odd values of $n$, for which $e^{-in\pi}=\pm 1$, respectively. After some additional rearrangement of the terms, this reads
\begin{linenomath*}
\begin{align}
    \ket{\psi_f} &= \dfrac{1}{2} \sum_{n\;\text{even}} c_n e^{i n (\phi\pm\pi/2)} \left( 2 e^{i\theta} \ket{e,n} + \e\sqrt{n} \left( e^{-i (\phi \mp\pi/2)} - 1 \right) (\ket{e,n-1}-e^{-i\theta}\ket{g,n-1}) \right. \notag \\
    &\phantom{\simeq\dfrac{1}{\sqrt{2}}} \left. +e^{i\theta} \e^\ast \sqrt{n+1} \left(1-e^{i (\phi\pm\pi/2)} \right) (\ket{g,n+1}+e^{i \theta}\ket{e,n+1})  \right) + \notag\\
    &\phantom{\simeq} + \dfrac{1}{2} \sum_{n\;\text{odd}} c_n e^{i n (\phi\pm\pi/2)} \left( 2\ket{g,n} - \e\sqrt{n} \left( e^{-i (\phi \mp\pi/2)} + 1 \right) (\ket{e,n-1}-e^{-i\theta}\ket{g,n-1}) - \right. \notag \\
    &\phantom{\simeq\dfrac{1}{\sqrt{2}}} \left. - e^{i\theta} \e^\ast \sqrt{n+1} \left(1+e^{i (\phi\pm\pi/2)}\right) (\ket{g,n+1}+e^{i \theta}\ket{e,n+1})  \right) + O(\epsilon^2)\;.
\end{align}
\end{linenomath*}
Finally, a measurement of the $\sz$ operator on the qubit state gives
\begin{linenomath*}
\begin{align}
    \langle\sz\rangle_{\theta} &= \bra{\psi_f}\sz \otimes \sum_m\ket{m}\bra{m} \ket{\psi_f}  \notag\\
    &= \sum_{n\;\text{even}} |c_n|^2 - \sum_{n\;\text{odd}} |c_n|^2 \notag\\
    &\phantom{=} + \dfrac{1}{2} \sum_{n\;\text{even}} \left[ c_n e^{in(\phi\pm\pi/2)}\Big(-c_{n+1}^\ast e^{-i(n+1)(\phi\pm\pi/2)}e^{i\theta}\e^\ast\sqrt{n+1}(e^{i(\phi\mp\pi/2)}+1) 
    \right. \notag\\
    &\hspace{45mm}\left.
    - c_{n-1}^\ast e^{-i(n-1)(\phi\pm\pi/2)}e^{-i\theta}\e\sqrt{n}(e^{-i(\phi\pm\pi/2)}+1)\Big) + \text{c.c.} \right] \notag\\
    &\phantom{=} + \dfrac{1}{2} \sum_{n\;\text{odd}} \left[ c_n e^{in(\phi\pm\pi/2)}\Big(c_{n+1}^\ast e^{-i(n+1)(\phi\pm\pi/2)}e^{i\theta}\e^\ast\sqrt{n+1}(e^{i(\phi\mp\pi/2)}-1) 
    \right. \notag\\
    &\hspace{45mm}\left.
    + c_{n-1}^\ast e^{-i(n-1)(\phi\pm\pi/2)}e^{-i\theta}\e\sqrt{n}(e^{-i(\phi\pm\pi/2)}-1)\Big)+ \text{c.c.} \right]  \notag\\
    &\phantom{=} + O(\e^2)  \;\\
    &= \sum_{n\;\text{even}} |c_n|^2 - \sum_{n\;\text{odd}} |c_n|^2 \notag\\
    &\phantom{=}+ \sum_{n\;\text{odd}} \textrm{Re}\Big(c_n c_{n+1}^\ast e^{-i(\phi\pm\pi/2)}e^{i\theta}(e^{i(\phi\mp\pi/2)}-1)\e^\ast\sqrt{n+1}  \notag \\
    &\phantom{= +\sum_{n\;\text{odd}} \textrm{Re} Aa}- c_{n+1} c_n^\ast e^{i(\phi\pm\pi/2)}e^{-i\theta}(e^{-i(\phi\pm\pi/2)}+1)\e\sqrt{n+1} \notag \\
    &\phantom{= +\sum_{n\;\text{odd}} \textrm{Re} Aa}+ c_n c_{n-1}^\ast e^{i(\phi\pm\pi/2)}e^{-i\theta}(e^{-i(\phi\pm\pi/2)}-1)\e\sqrt{n} \notag \\
    & \phantom{= +\sum_{n\;\text{odd}} \textrm{Re} Aa} - c_{n-1} c_n^\ast e^{-i(\phi\pm\pi/2)}e^{i\theta}(e^{i(\phi\mp\pi/2)}+1)\e^\ast\sqrt{n} \Big) \notag\\ 
    &\phantom{=}+ \frac{\abs{\e^2}}{2} \left(2\textrm{sin}\abs{\phi}- \sum_{n\;\text{even}} |c_n|^2 + \sum_{n\;\text{odd}} |c_n|^2\right)\notag\\
    & \phantom{=}+\textrm{Re}\left(\e^2 e^{-2i\theta}\left(1+e^{2i\phi}\right)\left(\sum_{n\;\text{even}}c_n c_{n-2}^\ast \sqrt{n(n-1)} -\sum_{n\;\text{odd}}c_n c_{n-2}^\ast \sqrt{n(n-1)} \right)\right)\notag\\
    &  \phantom{=}+O(\e^3) \;
    \label{eq:rigatoni}
\end{align}
\end{linenomath*}
Here we have added back the final result for order $\e^2$. The order $\epsilon^0$ expression $\Pi \equiv\sum_{n\;\text{even}} |c_n|^2 - \sum_{n\;\text{odd}} |c_n|^2$ is the result expected from an ideal parity measurement. The higher order terms, on the other hand, depend on $\epsilon$ and represent deviations from the ideal parity measurement. To partially cancel out these terms, we can average measurements using different phases $\theta$ for the qubit drive pulses. For example, it's straightforward to see from Equation \ref{eq:rigatoni} that $\langle\sigma_z\rangle\vert_{\theta}+\langle\sigma_z\rangle\vert_{\theta+\pi}$ cancels terms to first order in $\epsilon$. Averaging over four phases $\pi/2$ apart also cancels out part of the second order term (second to last line of Equation \ref{eq:rigatoni}), leaving only the term $|\e|^2 (\sin\abs{\phi}-\Pi/2)$.

\textbf{Echo sequence:} For the experiment using the echo sequence presented in the main text, we can extend the above calculation and see that the results are analogous. We start from Equation~\ref{eq:spaghetti} but for an interaction time $t_0^\prime=t_0/2=\pi/(2\abs{\chi})$. We then undo the SW transformation and apply the echo pulse corresponding to a counterclockwise rotation $R=e^{-i \frac{\eta}{2}\Vec{u}.\Vec{\sigma}}$ by $\eta=\pi$ and with the same phase as the $\pi/2|_\theta$ pulses. This transforms $\ket{g,n} \rightarrow e^{i\theta} \ket{e,n}$ and $\ket{e,n} \rightarrow -e^{-i\theta} \ket{g,n}$. Finally, we evolve the state according to the dispersive Hamiltonian with $\Delta\rightarrow-\Delta$ for another time $t_0^\prime$ and apply the second $\pi/2|_\theta$ that concludes the echo sequence. As in the Ramsey case, the expectation value $\langle\sigma_z\rangle\vert_{\theta}$ contains terms of order $\epsilon$ and higher that give a deviation form the ideal parity measurement. However, when averaging over measurements taken with four different $\theta$'s separated by $\pi/2$, we again see that all first order terms and some second order terms are canceled out. This approach results in an ideal parity measurement up to order $|\epsilon|^2$ that is also robust against low frequency qubit frequency flucuations, thanks to the echo pulse.
\begin{figure}[tbhp]
\centering
\includegraphics[width=16cm]{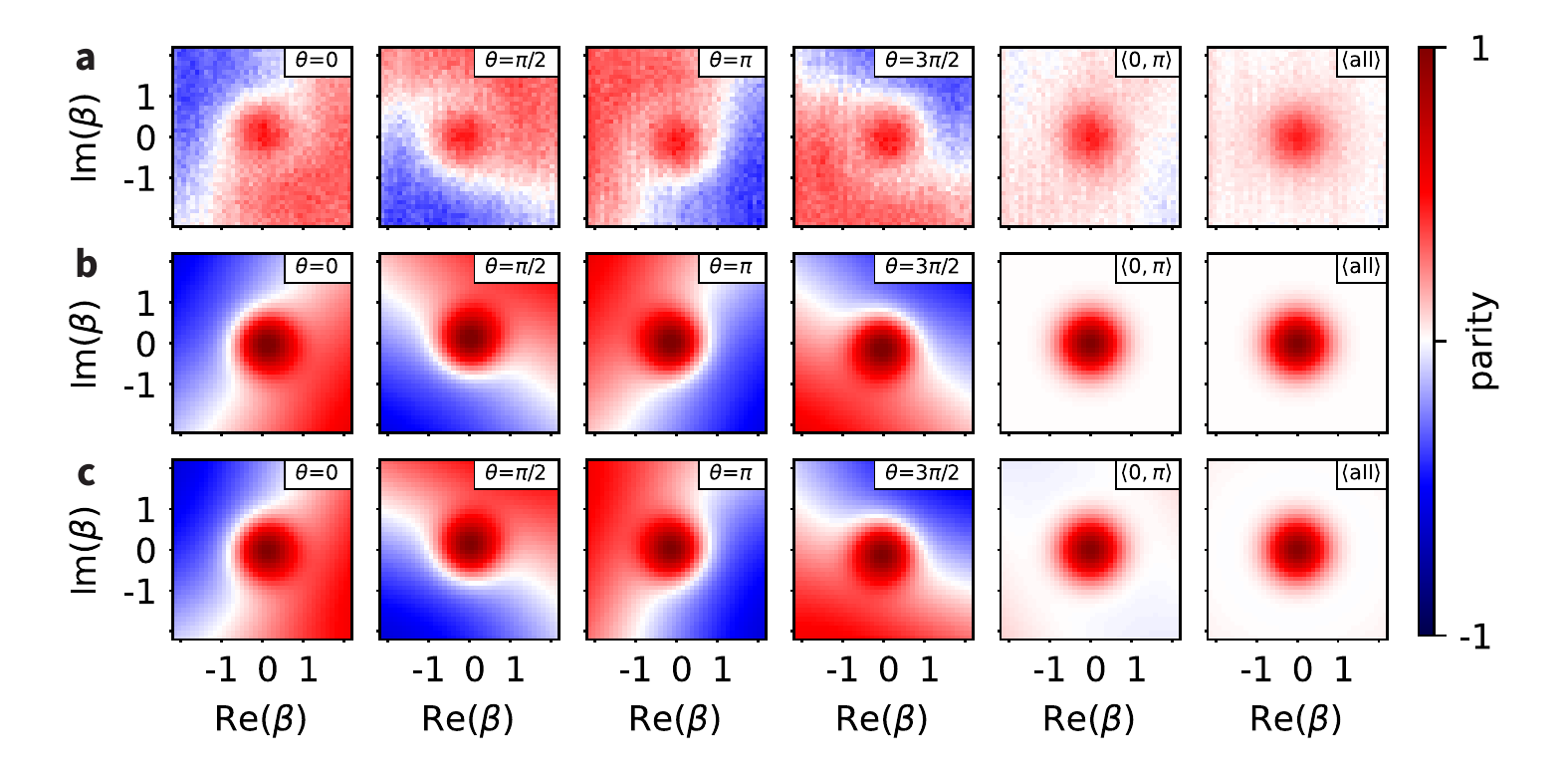}
\caption{
\textbf{Four phase averaging.} Comparison between measured \textbf{(a)}, analytically calculated \textbf{(b)}, and simulated \textbf{(c)} Wigner functions for four different initial phases $\theta$ of the qubit drive pulses (left four columns). The two rightmost columns illustrate the result of averaging over two phases, $(\langle\sigma_z\rangle\vert_{0}+\langle\sigma_z\rangle\vert_{\pi})/2$ or four phases $(\langle\sigma_z\rangle\vert_{0}+\langle\sigma_z\rangle\vert_{\pi/2}+\langle\sigma_z\rangle\vert_{\pi}+\langle\sigma_z\rangle\vert_{3\pi/2})/4$. Row \textbf{a} shows Wigner functions measured using the echo sequence shown in Figure \ref{fig4}a of the main text. Row \textbf{b} shows the analytical result in Equation \ref{eq:rigatoni} up to first order in $\epsilon$. Row \textbf{c} shows results of simulating the full Hamiltonian of the system (Equation \ref{eq:fullH}).  Rows \textbf{b} and \textbf{c} use a bare qubit-phonon detuning of $\Delta = -\unit{1.87}{MHz} - \chi/2 = -\unit{1.835}{MHz}$, and an interaction time of $\unit{7.04}{\mu s}$ was chosen so that the analytical and simulated results match the experimental ones. Both correspond to the experimental parameters to within uncertainties, taking into account qubit frequency fluctuations and finite ramp-up times of the Stark shift. Rows \textbf{b} and \textbf{c} also do not include the effects of decoherence.
\label{figS:wigner_phases} }
\end{figure}
In Figure~\ref{figS:wigner_phases}, we illustrate this approach for mitigating the non-idealities of the Wigner function measurement by comparing the experimental results to the analytical ones presented above, along with the results of simulating the full Hamiltonian of the system given by Equation \ref{eq:fullH}. 
% figure S4: 
\begin{figure}[tbhp]
\centering
\includegraphics[width=8cm]{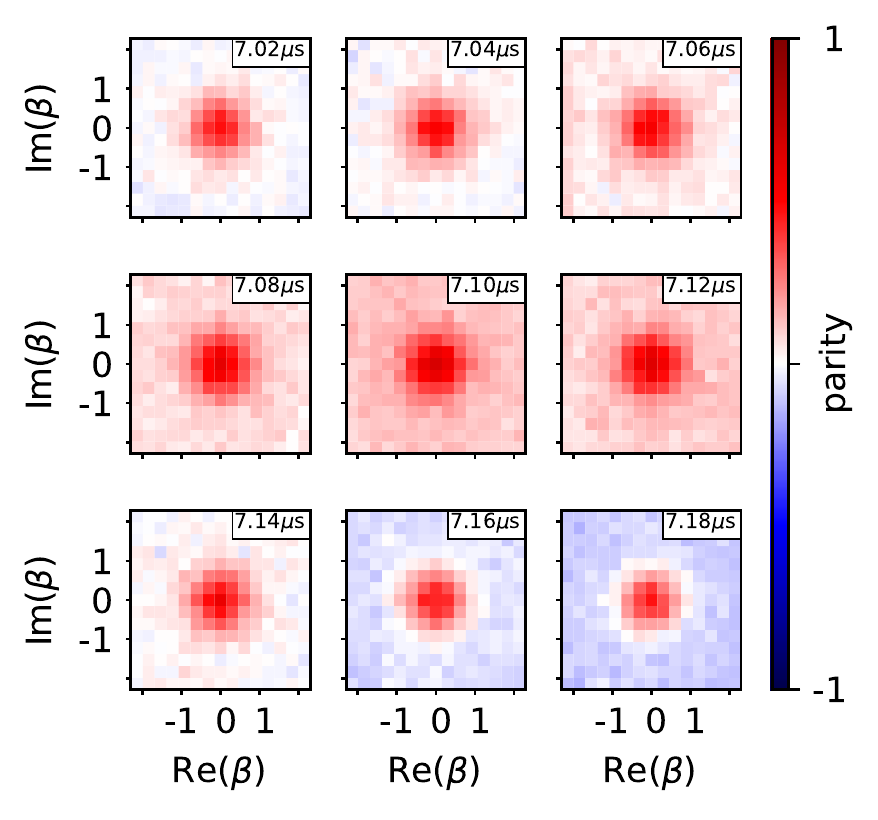}
\caption{
\textbf{Dependence of Wigner function on interaction time.}
Measured Wigner functions for different interaction times with the phonon in $\ket{0}$, with echo sequence and averaging over four values of $\theta$. The offset level of the Wigner function changes with the interaction time following a periodic pattern with a period of $\sim \unit{0.23}{\mu s}$. 
\label{figS:wigner_time} }
\end{figure} 

\textbf{Choice of interaction time.} \label{supp:t} As mentioned above, after averaging over data taken with four different values of $\theta$, we expect the remaining dominant effect of finite $\e$ to be a slight reduction in the Wigner function contrast, along with a constant offset given by $|\e|^2 \sin\abs{\phi}$. 
By varying the interaction time $t$, the value of this offset should oscillate with a frequency of $\Dpr \gg \abs{\chi}$. 
This means that we can choose $t$ such that the offset is zero without significantly affecting the parity measurement. 
In Figure \ref{figS:wigner_time}, we plot the measured Wigner functions for several values of $t$ that are all close to $\pi/\abs{\chi}$. 
We find that indeed there is a constant offset that depends on $t$, but the frequency of oscillations is $\sim 2\Dpr$ rather than $\Dpr$ as expected. 
We have checked that a master equation simulation using the full JC Hamiltonian including decoherence, finite pulse lengths, and the LG-10 mode agree well with our analytical results. 
Therefore, the origin of this discrepancy is unclear at the moment.
Nevertheless, we can still employ the strategy of choosing an interaction time that minimizes the offset. 
For the Wigner functions shown in Figure \ref{fig4}b, we use an interaction time of $\unit{7.05}{\mu s}$.
\newpage
\subsection{Dispersive shift measurements and predictions.}\label{supp:chi}
\begin{figure}
\centering
\includegraphics[width=8cm]{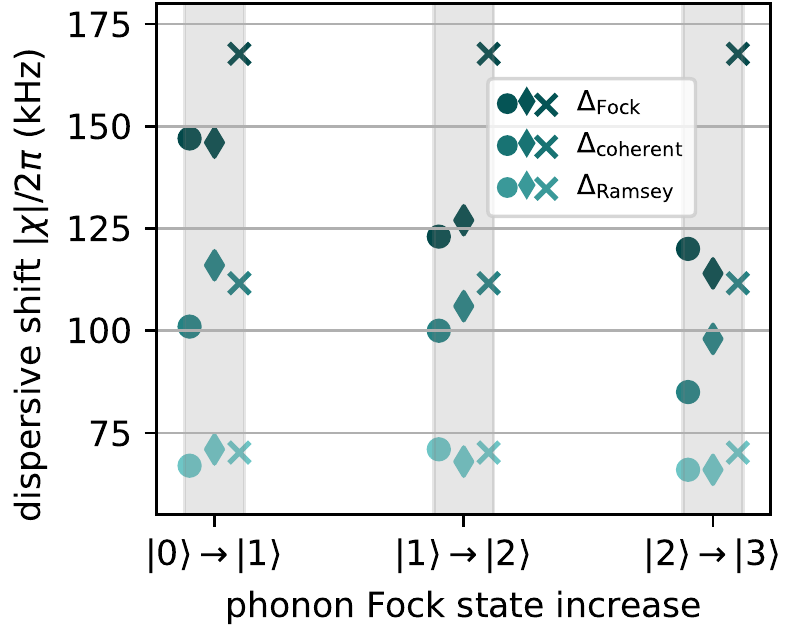}
\caption{
\textbf{Dispersive shifts for different detunings.} Measured ($\circ$), numerical ($\lozenge$), and analytical ($\times$) absolute values of the qubit frequency shift due to an increase in phonon population by one quantum. $\circ$ and $\times$ markers are shifted horizontally to increase visibility. Colors indicate the qubit-phonon detuning at which the measurement was taken. Refer to Table \ref{tab:device} for numerical values of the detunings.
\label{figS:chis} }
\end{figure}

Here, we discuss in detail how the qubit frequency shifts depends on the phonon population, taking into account departures from the ideal dispersive Hamiltonian. 
We first extract the qubit frequencies corresponding to different phonon numbers from two spectroscopy measurements, one with the qubit at $\Delta_{\mr{coherent}}$ (Figure \ref{fig2}a, displacement amplitude = 0.4) and another at $\Delta_{\mr{Fock}}$ (Figure \ref{fig3}a, M = 3), as well as from fitting the frequency of time-domain oscillations in the qubit population with the qubit at $\Delta_{\mr{Ramsey}}$ (see Figure \ref{fig3}b of the main text). We then compute the qubit frequency difference between $n$ and $n+1$ phonons. 
In  Figure \ref{figS:chis}, these measured values ($\circ$) are compared with results from numerical diagonalization of the JC Hamiltonian ($\lozenge$) and with the analytical result of Equation \ref{eq:chi} in the main text ($\times$). Unlike the constant shift of the qubit frequency by $\chi$ for each additional phonon predicted by the dispersive Hamiltonian of Equation 1 in the main text, we find that both the measured and calculated frequency shifts per phonon decreases in absolute value with phonon number.
Furthermore, we observe this decrease to be more significant for smaller qubit-phonon detunings. 

We attribute both these effects to the contribution of terms of higher order in $g/\Delta$ arising from the Schrieffer-Wolff transformation. Those terms are negligible in the strong dispersive regime where $\abs{g/\Delta} \ll 1$. 
A higher initial Fock state $n$ increases the effective coupling strength between qubit and phonon to $\sqrt{n}g$. 
Thus, both a smaller $\Delta$ and larger initial $n$ increase the ratio $\abs{g/\Delta}$, requiring higher order terms to be taken into account. 
We note that for the detuning $\Delta_\mr{Ramsey}$ where we performed parity measurements (Figure \ref{fig3}b) and Wigner tomography (Figure \ref{fig4}) the change in frequency shift with increasing initial Fock state is small, indicating operation well within the dispersive regime. 

% figure chi values: 

\end{document}